\RequirePackage{ifpdf}
\documentclass[preprint,notoc]{JHEP3}
\usepackage{epsfig}
\usepackage{slashed}

\def\sr2{\sqrt{2}}  
 \def\to{\rightarrow} 
\def\bi{\begin{itemize}} \def\ei{\end{itemize}} 
\def\c1p{C1^\prime} \def\ta{\tilde a}  
\def\th{\tilde h} \def\thetab{\bar\theta}  
\def\tu{\tilde u}  \def\ta{\tilde a} 
\def\tb{\tilde b}   
  \def\tst{\tilde t} 
 \def\tg{\tilde g}  \def\tell{\tilde\ell}
\def\tq{\tilde q}  \def\tw{\widetilde W} \def\tz{\widetilde
Z} 
  \def\alt{\lesssim}
\def\agt{\gtrsim} \def\be{\begin{equation}} \def\ee{\end{equation}}
\def\bea{\begin{eqnarray}} \def\eea{\end{eqnarray}}

\preprint{\vbox{OU-HEP-150228}}

\title{Mixed axion-wino dark matter } \author{Kyu Jung Bae$^{a}$, Howard
Baer$^{a}$, Andre Lessa$^{b}$ and Hasan Serce$^{a}$\\
$^a$Dept.\ of Physics and Astronomy, University of Oklahoma, Norman, OK 73019,
USA\\
$^b$ Instituto de F\'isica, Universidade de S\~ao Paulo, S\~ao Paulo - SP,
Brazil\\
E-mail: \email{bae@nhn.ou.edu}, \email{baer@nhn.ou.edu},
\email{lessa@if.usp.br}, \email{serce@ou.edu} }

\abstract{ A variety of supersymmetric models give rise to a split mass spectrum
characterized by  very heavy scalars but sub-TeV gauginos, usually with a
wino-like LSP. Such models predict a thermally-produced underabundance of
wino-like WIMP dark matter so that non-thermal DM production mechanisms are
necessary.
We examine the case where theories with a wino-like LSP are augmented by a
Peccei-Quinn sector including an axion-axino-saxion supermultiplet in either the
SUSY KSVZ or SUSY DFSZ models and with/without saxion decays to axions/axinos.
We show allowed ranges of PQ breaking scale $f_a$ for various cases which are
generated by solving the necessary coupled Boltzmann equations.
We also present results for a model with radiatively-driven naturalness
but with a wino-like LSP. }

\begin{document}

\section{Introduction}
\label{sec:intro}

Supersymmetric models with anomaly-mediated SUSY breaking~\cite{amsb} (AMSB)
provided a strong {\it raison d'\^{e}tre} for considering the case of a
wino-like lightest SUSY particle, or LSP. Originally, such models were built
with a ``sequestered''-- rather than a hidden-- SUSY breaking sector.
The sequestered sector could be located on a brane which was separated from the
visible sector brane in an  extra dimensional space-time.
In such a case, tree level supergravity contributions to soft SUSY breaking
terms were absent and the dominant contribution to soft terms came from the
superconformal anomaly.
Since the soft terms were all of order $m_{\rm soft}\sim m_{3/2}/(16\pi^2)$,
then values of gravitino mass $m_{3/2}\sim 30-100$ TeV were required to generate
a weak-scale sparticle mass spectrum. The weak-scale gaugino masses were
expected to occur in the ratio $M_1:M_2:M_3\sim 3:1:8$, resulting in a wino-like
LSP as the dark matter candidate.
The thermally-produced relic density of a wino-like LSP is typically~\cite{Wells:2003tf, Arkani2006} 
\be 
\Omega_{\tw}^{\rm TP}h^2\sim 0.12\left(M_2/2.5\ {\rm TeV}\right)^2 .
\ee 
The measured dark matter abundance $\Omega_{\rm CDM}h^2=0.12$ is then
saturated for a wino of mass $m_{\tw_1}\simeq M_2 \sim 2.5$ TeV.
For lighter winos,  non-thermal production mechanisms such as WIMP production
from moduli decay were invoked~\cite{mr}.

While the simplest AMSB models provided solutions to the SUSY flavor, CP and
gravitino problems, they retain the problem of predicting tachyonic slepton
masses.
More recently, they may have fallen into disfavor due to the
discovery~\cite{atlas_h,cms_h} of the Higgs boson with mass $m_h =125.5\pm 0.5$
GeV.
In the minimal AMSB model, this value of Higgs mass requires $m_{3/2}\sim 1000$
TeV so that the sparticle mass spectrum lies in the multi-TeV region which seems
to seriously compromise even the most conservative measures of
naturalness~\cite{dm125,seige}.

Even well-before the Higgs discovery, related models with a wino-like LSP were
emerging.
These include
\begin{itemize}
\item PeV SUSY~\cite{Wells:2003tf,pev},
\item split SUSY~\cite{ArkaniHamed:2004fb,Giudice:2004tc,ArkaniHamed:2004yi},
\item G2MSSM~\cite{g2mssm},
\item models with strong moduli stabilization~\cite{Kallosh:2004yh},
\item pure gravity mediation~\cite{pgm,pgm_mu} and
\item spread SUSY~\cite{Hall:2011jd,Hall:2012zp}.
\end{itemize}

These models differ from the original mAMSB model in that they predict a split spectrum 
with scalars ranging from 25 TeV all the way to $\sim 10^{8}$ TeV-- well beyond the
reach of collider experiments. 
In contrast, the gauginos typically lie in the $0.1-3$ TeV region so that the lower
range of values would be accessible to LHC searches. 
In most of these models, the gauginos adopt either the AMSB-form~\cite{pev,pgm} 
or a mixed anomaly plus loop contribution form~\cite{g2mssm,Hall:2011jd,Hall:2012zp} 
which also typically gives rise to a wino-like LSP.
The SUSY $\mu$ parameter is variable between these several models
and may be as small as $\sim 1$ TeV~\cite{g2mssm,pgm_mu} or as high as hundreds of TeV~\cite{pgm}.
While the predicted thermal abundance of wino-like WIMPs saturates the measured
value for a wino mass of $\sim 2.5$ TeV (so the gaugino spectrum would be well beyond 
reach of LHC), for lower $M_2$ values a thermal underabundance of
WIMPs is expected and some non-thermal DM production mechanism is 
needed. Usually, this has involved some form of moduli production 
and decay~\cite{mr,gondolo,g2DM,Allahverdi:2012gk} (for recent reviews, see Ref's~\cite{Baer:2014eja,Kane:2015qea}).

In the present paper, we instead look at non-thermal wino production from the 
Peccei-Quinn (PQ) sector.\footnote{An earlier look at non-thermal production of winos
in AMSB models was given in Ref.~\cite{Baer:2010kd}.} 
By invoking a PQ sector in supersymmetric models~\cite{Kim:1983ia} 
the axion supermultiplet also contains an $R$-parity-even spin-0 saxion $s$ 
and an $R$-parity-odd spin-$1/2$ axino $\ta$. 
This approach has several advantages:
\bi
\item a PQ sector seems necessary to solve the strong CP problem in the QCD 
sector~\cite{pqww,ksvz,dfsz},
\item invoking PQ charges for Higgs multiplets offers a means to forbid the
appearance of a Planck scale $\mu$ term while re-generating a weak-scale $\mu$ 
term (solution to the SUSY $\mu$ problem)~\cite{kn},
\item while the presence of the PQ sector can act to augment the wino abundance--
for instance by axino and/or saxion decays-- the axion abundance can always be 
adjusted to make up any remaining DM abundance which may be needed. 
\ei

To explore this situation, we will adopt a benchmark model which encapsulates 
the dark matter physics expected in the above list of models. This benchmark point--
labelled as CSB for ``charged SUSY breaking''~\cite{pev}-- 
contains scalar masses around the 72 TeV region while gauginos lie in the $0.2-2$ TeV
range. The thermally-produced WIMP abundance is predicted to be
$\Omega_{\tw}h^2\sim 0.002$-- a factor $\sim 60$ below the measured value.
Such a low thermal WIMP abundance requires additional dark matter production
mechanisms to match experiment. In the case presented here, the dark matter
is actually composed of both WIMPs and axions. While WIMPs can be produced thermally, 
they can also be produced via axino, saxion and gravitino production and 
decay in the early universe. In addition, saxions produced via coherent 
oscillations (CO) can inject late-time entropy into the early universe, thus diluting any relics already present.
Axions can be produced as usual via CO~\cite{vacmis,vg}, 
but can also be produced thermally and via saxion decay.

While the models listed above are motivated by a variety of theoretical 
and phenomenological considerations, we note that collectively the entire set
is highly fine-tuned in the electroweak sector, since the weak scale
values of $m_{H_u}^2$ and $\mu^2$ would have to be adjusted to very high precision to gain a $Z$ mass of just 
$91.2$ GeV. Thus, for contrast, we also examine a SUSY model with radiatively-driven
naturalness~\cite{rns} but with a wino-like LSP~\cite{bbhmpt} with fine-tuning at just the 10\% level
(labelled as RNSw).

In Sec.~\ref{sec:models}, we briefly review a variety of models with split spectra and
a wino-like LSP. We also present a SUSY model with radiatively-driven naturalness 
and a wino-like LSP for comparison.  In Sec.~\ref{sec:boltz}, we briefly review our coupled-Boltzmann
equation evaluation of mixed axion/wino dark matter 
(more details can be found in Ref.~\cite{Bae:2014rfa}).
In Sec.~\ref{sec:results}, we present the results of our coupled Boltzmann
computation of the mixed axion/wino dark matter abundance in the CSB and RNSw
benchmark models. In Sec.~\ref{sec:lines}, we expand our two benchmark points to model lines
to examine how our results depend on the SUSY mass spectrum. 
Our overall conclusions and a summary plot are given in Sec.~\ref{sec:conclude}.

\section{Survey of some models with a wino-like LSP}
\label{sec:models}

\subsection{PeV SUSY} 

In Ref.~\cite{Wells:2003tf,pev}, it is argued that the PeV scale 
(with $m({\rm scalars})\sim m_{3/2}\sim 1$ PeV$=$1000 TeV) is motivated by considerations of 
wino dark matter and neutrino mass while providing a decoupling 
solution~\cite{dine} to the SUSY flavor, CP, proton decay and gravitino/moduli problems.
This model invoked ``charged SUSY breaking'' (CSB) where the hidden sector superfield $S$
is charged under some unspecified symmetry. 
In such a case, the scalars gain masses via
\be
\int d^2\theta d^2\thetab \frac{S^\dagger S}{M_P^2}\Phi_i^\dagger \Phi_i\Rightarrow
\frac{F_S^\dagger F_S}{M_P^2}\phi_i^*\phi_i
\ee
while gaugino masses, usually obtained via gravity-mediation as
\be
\int d^2\theta\frac{S}{M_P}WW\Rightarrow \frac{F_s}{M_P}\lambda\lambda ,
\ee
are now forbidden. 
Then the dominant contribution to gaugino masses comes from AMSB:
\bea
M_1&=& \frac{33}{5}\frac{g_1^2}{16\pi^2}m_{3/2}\sim m_{3/2}/120,\\
M_2&=& \frac{g_2^2}{16\pi^2}m_{3/2}\sim m_{3/2}/360,\\
M_3&=& -3\frac{g_3^2}{16\pi^2}m_{3/2}\sim -m_{3/2}/40 .\\
\eea
Saturating the measured dark matter abundance with thermally-produced winos 
requires $m_{\tw}\sim M_2 \sim 2.5$ TeV which in turn requires the gravitino and scalar masses
to occur at the $\sim 1000$ TeV (or 1 PeV) level.
The author remains agnostic as to the magnitude of $\mu$, although $\mu\gg M_2$ is expected.

\subsection{Split SUSY}

In Split SUSY~\cite{ArkaniHamed:2004fb,Giudice:2004tc,ArkaniHamed:2004yi,
Pierce:2004mk,Arvanitaki:2004eu}, 
SUSY is still required for gauge coupling unification and for 
a dark matter candidate, but naturalness is eschewed in favor of a 
multi-verse solution to the gauge hierarchy problem. 
In such a case, matter scalars can exist with masses typically at some
intermediate scale $m_{\tq,\tell}\sim 10^{8}$ TeV while SUSY fermions
(gauginos and higgsinos) are protected by chiral symmetry and can be much lighter.
Split SUSY can be realized under charged SUSY breaking as in PeV-SUSY or via
Scherk-Schwartz SUSY breaking in extra dimensions~\cite{ArkaniHamed:2004fb}.
Here, one might expect
\be
m({\rm gauginos})\sim m({\rm higgsinos})\ll m({\rm scalars})
\ee
where the authors remain agnostic concerning whether the wino or bino might be 
lighter. 
Typically, binos should overproduce dark matter so that a wino/higgsino
admixture might be expected.

\subsection{G2MSSM} 

In string/M-theory models which are compactified on
a manifold of $G_2$ holonomy~\cite{g2mssm}, one expects a gravitino mass
$m_{3/2}\sim 25-100$ TeV along with a cosmologically relevant 
moduli field with similar mass~\cite{Acharya:2010af}. The matter scalar masses are of 
order $\sim m_{3/2}$ but gaugino masses can be much lighter. 
Typically, a wino LSP is to be preferred~\cite{g2DM}. 
The superpotential $\mu$ term is generated with value $\sim 1$ TeV so that 
these models tend to be more electroweak-natural than split SUSY.

\subsection{Models with strong moduli stabilization (Kallosh-Linde or KL)}

In string theory, an outstanding problem exists in the need for vacuum stabilization of moduli
fields. In the KKLT construction~\cite{kklt}, one constructs a stable supersymmetric 
anti-deSitter vacuum, but then uplifts to a deSitter vacuum via SUSY breaking.
In KKLT, the volume modulus mass $m_\sigma$ is expected to be comparable to the
gravitino mass $m_{3/2}$.
These models give rise to soft SUSY breaking terms characterized by comparable 
moduli- and anomaly-mediated contributions~\cite{Choi:2004sx}.
However, these models suffer from vacuum destabilization during inflation 
unless the Hubble constant $H< m_{3/2}$. Such inflationary models, while possible, are
often unwieldy and inelegant~\cite{Kallosh:2004yh}. 

An alternative approach known as strong vacuum stabilization
invokes instead a racetrack superpotential for the volume modulus, leading to
a far heavier modulus mass $m_{\sigma}\sim 10^{15}$ GeV and allowing for 
vacuum stability in models of chaotic inflation~\cite{Kallosh:2004yh}. 
In this Kallosh-Linde (KL) case~\cite{Linde:2011ja}, 
the soft SUSY breaking scalar masses are comparable to $m_{3/2}$, but the
gaugino and trilinear soft terms are suppressed by a factor of $m_{3/2}/m_\sigma$. 
The dominant contribution to gaugino masses comes from anomaly-mediation. Requiring a wino LSP
without too much relic density then fixes $m_{3/2}\alt 1000$ TeV.
Thus, one gains a model of split SUSY with PeV-scale scalar masses but with TeV-scale
gauginos with an AMSB mass pattern. The $\mu$ parameter is also expected
to be $\sim m_{3/2}$~\cite{dlmmo} so a high degree of electroweak fine-tuning is needed.

\subsection{Pure gravity-mediation}

In pure gravity mediation (PGM) models~\cite{pgm}, it is assumed that matter scalar masses
are developed at tree level and so have masses $m_{\tq,\tell,H}\sim m_{3/2}\sim 
1000$ TeV while gaugino masses are suppressed since no SUSY breaking 
fields are assumed to be singlets under any symmetries. The gaugino masses
arise via anomaly mediation so the wino is expected to be the LSP. 
The $\mu$ term and SUSY breaking bilinear $B$
are also expected to be at the $m_{3/2}$ scale leading to
\be
m_{\tg,\tb,\tw}\ll m_{\tq,\tell,H,\th}\sim 100\ {\rm TeV}\ \ \ \ ({\rm PGM})
\ee
although a recent incarnation also allows for light higgsinos~\cite{pgm_mu}.

\subsection{Spread SUSY} 

In Spread SUSY~\cite{Hall:2011jd,Hall:2012zp}, 
additional spatial dimensions are 
assumed so that the 4-d reduced Planck scale $M_P$ is enhanced by a volume factor
over the fundamental scale $M_*$. Then, if the hidden sector SUSY breaking field
$X$ is charged under some symmetry, gaugino masses are generated only via anomaly-
mediation while scalar masses are generated via gravity-mediation. One expects a mildly
split-- or spread-- SUSY spectrum characterized by
\be
m_{\tw,\tb,\tg}\ll m_{3/2}\sim m_{\th}\ll m_{\tq,\tell,H}
\ee
where the wino is the LSP with sub-TeV masses and the matter scalar masses may
lie in the $10^2-10^3$ TeV range while the higgsinos are intermediate between these two.

\subsection{Natural SUSY with wino-like LSP}

In SUSY with radiatively-driven naturalness~\cite{rns}, the
$W,Z,h$ mass scale arises naturally due to a supersymmetric $\mu$ 
parameter with $\mu\sim 100-300$ GeV (the closer to $m_Z$ the better)
while $m_{H_u}^2$ is driven radiatively to small rather than large values.
The TeV-scale top squark masses are highly mixed which uplifts
$m_h$ to $\sim 125$ GeV whilst suppressing radiative corrections
to the scalar potential which influence the values of $m_{h,Z}$.
While one expects a higgsino-like LSP under conditions of gaugino mass
universality, models with non-universal gaugino masses allow for 
a bino-like or wino-like LSP without sacrificing naturalness~\cite{bbhmpt}.
Mixed axion-higgsino dark matter has been previously calculated
in Ref's~\cite{Bae:2013qr,dfsz1,Bae:2014rfa} while the mainly bino-like LSP 
case is largely excluded due to overproduction 
of WIMPs~\cite{Bae:2013qr,Bae:2014rfa}.
Here, we consider the wino-like LSP case which typically yields
a thermally-produced wino abundance of $\Omega_{\tw}h^2\sim 0.001$ 
for winos with $m_{\tw}\sim 100-200$ GeV 
(at least an order-of-magnitude lower than expectations for a similarly massive
higgsino LSP). 

\subsection{Two benchmark points}

In order to compute the mixed axion/wino dark matter relic abundance 
in the SUSY axion models, we must specify both the PQ and the MSSM parameters. 
On the MSSM side, we adopt two SUSY benchmark models for illustration. 

The first has been listed as benchmark CSB since it occurs in the rather
simple and elegant charged SUSY breaking model of Ref.~\cite{pev}.
It is rather similar to the Kallosh-Linde~\cite{Kallosh:2004yh} 
benchmark from the study of Ref.~\cite{Baer:2013ula}. 
We take the CSB benchmark to be illustrative of the
large class of models with multi-TeV scalars but with sub-TeV gauginos
with a wino as LSP. The CSB benchmark model is listed in Table~\ref{tab:bm}.
We generate the SUSY model spectra with Isajet 7.83~\cite{isajet}.

Along with the CSB benchmark, we adopt a natural SUSY benchmark with a wino as
LSP. It is taken from Ref.~\cite{bbhmpt} and denoted as RNSw 
(radiatively-driven natural SUSY with a wino LSP).

%
\begin{table}\centering
\begin{tabular}{lcc}
\hline
 & CSB & RNSw   \\
\hline
$m_{0}$ & 72000 & 5000 \\
$M_1$ & 1320 & 700 \\
$M_2$ & 200 & 175 \\
$M_3$ & -600 & 700 \\
$A_0$ & 0 & -8000  \\
$\tan\beta$  & 10 & 10  \\
$\mu $ & 3000 & 200  \\
$m_A$ & 72000 & 1000 \\
$m_h$ & 126.0 & 124.3 \\
$m_{\tg}$ & 1924 & 1810 \\
$m_{\tu_L}$ & 71830 & 5101 \\
$m_{\tst_1}$ & 47760 & 1478 \\
$m_{\tz_2}$ & 635.9 & 211.8 \\
$m_{\tz_1}$ & 203.2 & 114.2 \\
\hline
$\Delta_{EW}$ & 22830 & 10.78 \\
$\Omega^{\rm std}_{\tz_1} h^2$ & 0.0020 & 0.0015 \\
$\sigma^{\rm SI}(\tz_1 p)$ pb & $6.2\times 10^{-12}$ & $4.3\times 10^{-8}$ \\
$\sigma^{\rm SD}(\tz_1 p)$ pb & $1.4\times 10^{-8}$ & $9.0\times 10^{-4}$ \\
$\langle\sigma v\rangle |_{v=0}$ cm$^3$/s & $1.7\times 10^{-24}$ & $1.7\times 10^{-24}$ \\
\hline
\end{tabular}
\caption{Masses and parameters in~GeV units for two benchmark points
computed with Isajet 7.83 and using $m_t=173.2$ GeV.
}
\label{tab:bm}
\end{table}

\section{Brief review of coupled Boltzmann calculation}
\label{sec:boltz}

To accurately estimate the mixed axion/neutralino dark matter production rate
in the early universe, it is necessary to evaluate the coupled Boltzmann
equations which track dark matter number densities and energy densities in an intertwined
manner. The exact equations used are presented in Ref.~\cite{Bae:2014rfa} 
and will not be repeated here.
In our calculations, we use a combination of 
IsaReD~\cite{isared} and micrOMEGAs~\cite{Belanger:2013oya} for the evaluation of 
the wino annihilation cross-section ($\langle\sigma v\rangle$).

The relevant equations track the following number and energy densities:
\begin{enumerate}
\item neutralino densities including thermal production and production via 
decays of heavier partices ({\it e.g.} axinos, saxions and gravitinos) followed
by possible subsequent re-annihilation,
\item thermally-produced axinos along with axino production via heavy particle decays
and diminution of axinos due to their decays,
\item thermally produced saxions along with diminution via their decays,
\item thermally-produced gravitinos~\cite{gravprod} along with gravitino decay~\cite{gravdecay},
\item thermally-produced axions along with axion production via heavy particle decays,
\item axion production via coherent oscillations (CO) and
\item saxion production via CO along with saxion decays.
\item Along with these, we track the radiation density of  SM particles. 
\end{enumerate}
For thermal saxion and axion production, it is reasonable to expect
annihilation/production rates which are similar to axinos.

The above eight components result in 16 coupled Boltzmann equations: one for
the number density and one for the energy density of each component.
Together with the Friedmann equation $H=\sqrt{\rho_T/3M_P^2}$ (where $\rho_T$ is the 
energy density summed over all contributions and $M_P$ is the reduced Planck scale)
the Boltzmann equations form a closed system which may be solved numerically. 

For the SUSY KSVZ model, the various axino ($\ta\to g\tg$, $Z\tz_i$ and $\gamma\tz_i$) 
and saxion branching fractions ($s\to gg$, $\tg\tg$) can be found in Ref's~\cite{az1,bkls}.
In addition, the model-dependent decays $s\to aa$, $\ta\ta$ are effectively parameterized~\cite{cl}
by $\xi =\sum_i q_i^3v_i^2/v_{PQ}^2$ where $q_i$ are the 
charge assignments of PQ multiplets and $v_i$ are their vevs after PQ symmetry breaking
and $v_{PQ}=\sqrt{\sum_i v_i^2q_i^2}$. We will take $\xi =0$ or 1 which effectively turns
off or on saxion decays to axinos/axions~\cite{Bae:2013qr}. 
The decay $s\to \ta\ta$ augments the LSP
abundance whilst the decay $s\to aa$ leads to dark radiation parameterized by the
effective number of extra neutrinos present in the early universe $\Delta N_{\rm eff}$. 
The Planck Collaboration reported $N_{\rm eff}=3.52^{+0.48}_{-0.45}$ by the combined data 
($95\%$; Planck+WP+highL+$H_{0}$+BAO)~\cite{planck}.\footnote{As this paper was being finalized, 
this value was updated~\cite{Planck:2015xua} to $N_{eff}=3.15\pm 0.23$.} 
We require the upper bound $\Delta N_{\rm eff}<1$ as a reference value lest too much dark radiation 
is produced. Excluded points with $\Delta N_{\rm eff}>1$ are color-coded in our results.

For the SUSY DFSZ model, axino and saxion decay rates are very different from the KSVZ case.
While in the KSVZ model axino and saxion decay primarily to gauge bosons
and gauginos, in SUSY DFSZ then typically $\ta\to \tz_i\phi$ (where $\phi=h,H,A$),
$\tz_i Z$, $\tw_j W$ and $\tw_j^\mp H^\pm$ and $s\to $ pairs of Higgs bosons,
vector bosons and electroweak-ino pairs.
Complete formulae for the DFSZ decay rates are found in Ref.~\cite{dfsz1}.

The thermal production rates for SUSY KSVZ (which are proportional to $T_R$) 
are found in Ref's~\cite{ksvzprod} while thermal production rates for SUSY DFSZ (which are mostly independent of $T_R$)
are obtained from Ref's~\cite{bci}.
We include production of particles via both decays and inverse decays~\cite{Bae:2014rfa}: 
the latter effects are important in SUSY DFSZ where saxions and axinos are maximally produced at 
$T\sim m(\rm particle)$ which leads to a freeze-in effect~\cite{freezein} which manifests itself 
essentially as delayed saxion/axino decays.

\begin{figure}
\begin{center}
\includegraphics[height=9cm]{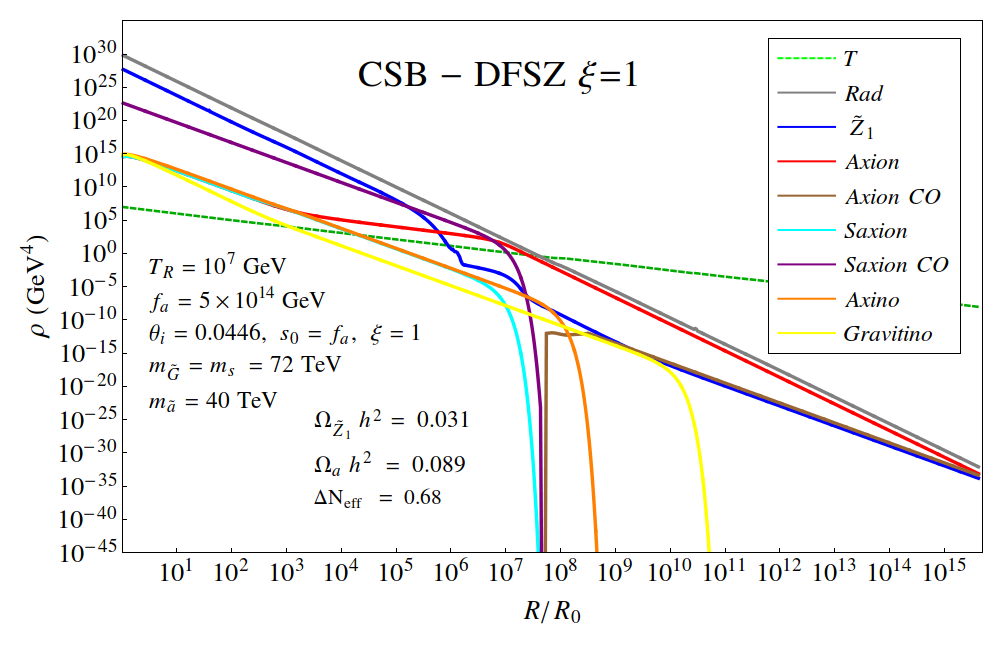}
\caption{Evolution of various energy densities vs. scale factor $R/R_0$
for the CSB benchmark case in SUSY DFSZ with $\xi=1$ and other parameters as indicated
in the figure.
\label{fig:KL}}
\end{center}
\end{figure}

An example of the evolution of various energy densities $\rho_i$ vs.
the cosmic scale factor $R/R_0$ is
shown in Fig.~\ref{fig:KL} for the SUSY DFSZ model. 
$R_0$ is taken to be the scale factor at the end of reheating ($T=T_R$). 
In the figure, $f_a=5\times10^{14}$ GeV while $m_s=m_{3/2}=72$ TeV for the CSB benchmark point. 
We also take $m_{\ta}=40$ TeV and $\xi =1$ so that saxion decay to
axions is turned on. At $R/R_0=1$, the universe is indeed radiation dominated
(gray curve) while including
a thermal population of WIMPs, saxions, axions, axinos and gravitinos. It also includes a
CO-component of saxions. As $R/R_0$ increases (decreasing temperature as denoted by the 
green dashed line),
the oscillating saxion field begins to decay-- mainly via $s\to aa$-- so that the
population of thermal/decay-produced axions (red curve) increases beyond its otherwise 
thermal trajectory around and below $R/R_0\sim 10^4$. The neutralino abundance (dark blue)
begins to freeze-out around $R/R_0\sim 10^5$, but then is augmented by decaying CO saxions 
and also by axinos (which decay slightly after saxions).
Decaying gravitinos add, but only marginally, to the neutralino abundance around
$R/R_0\sim 10^{10}$.
At $R/R_0\sim 10^8$, the axion mass turns on and the axion field begins to oscillate
as non-relativistic matter (brown curve).
Also, at $R/R_0\sim 10^9$, the neutralinos become non-relativistic. Together, the combined 
neutralino-axion CDM ultimately dominates the universe at around $R/R_0\sim 10^{16}$.
The ultimate dark matter density is composed of $\sim 25\%$ wino-like WIMPs and $\sim 75\%$
cold axions with a modest-but-not-yet-excluded 
contribution of relativistic axions ($\Delta N_{\rm eff}=0.68$) as dark
radiation.
%

%

%
%

\section{Mixed axion-wino dark matter}
\label{sec:results}

In the following subsections, we compute the
neutralino and axion relic abundances for the two benchmark points through
numerical integration of the Boltzmann equations as discussed in Sec.~\ref{sec:boltz}.
To gain more general results, we will scan over the PQ scale $f_a$ and the axino mass
which we take to be bounded by $m_{3/2}$:
\bea
 10^9 \mbox{ GeV} < & f_a & < 10^{16} \mbox{ GeV}, \nonumber \\
 0.4 \mbox{ TeV} < & m_{\ta} & < m_{3/2},
\label{eq:scan}
\eea
with $m_s$ fixed as $m_s=m_{3/2}$.
In many supergravity models, 
saxion mass is generated by the same operators as those for the MSSM scalars 
while axino mass is highly model dependent and can be much smaller than $m_{3/2}$~\cite{gy,cl}.
For this reason, we consider the above parameter range for our general analyses. 

For simplicity, we will fix the initial saxion field strength,
which sets the amplitude of coherent saxion oscillations, to $s_i=f_a$ ($\theta_s\equiv s_i/f_a =1$).
In addition-- for points which are DM-allowed ($\Omega_{\tz_1}h^2<0.12$) and obey
BBN and dark radiation constraints-- the initial axion mis-alignment
angle $\theta_i$ is set to the required value such that $\Omega_{\tz_1}h^2+\Omega_a h^2=0.12$.
In the SUSY DFSZ case, unlike the SUSY KSVZ model, the bulk of our results do
not depend strongly on the re-heat temperature ($T_R$) since the axion, axino and saxion TP rates are
largely independent of this quantity.
Nonetheless, the gravitino thermal abundance is proportional to $T_R$
and since gravitinos are long-lived they may affect BBN or WIMP abundance 
constraints if $T_R$ is sufficiently large.
In order to avoid the BBN constraints on gravitinos, we choose $T_R=10^7$ GeV,
which results in a sufficiently small (would-be) gravitino abundance. 
As a result, gravitinos typically do not contribute significantly to the neutralino
abundance, as discussed above.

For each of the CSB and RNSw benchmark points, we consider two different cases: $\xi=0$ 
(saxion decay to axions/axinos turned off) and $\xi=1$ (saxion decay to axions/axinos
turned on). We adopt a KSVZ model with SU(2)$_L$ singlet heavy quark states so that 
the axion superfield only has interactions with SU(3)$_c$ and U(1)$_Y$ gauge
superfields.
We discuss the case of SU(2)$_L$ doublet heavy quark states 
in Sec.~\ref{sec:conclude} for completeness.

\subsection{CSB benchmark in SUSY KSVZ}

\subsubsection{$\xi=0$ case}

\begin{figure}
\begin{center}
\includegraphics[height=6.9cm]{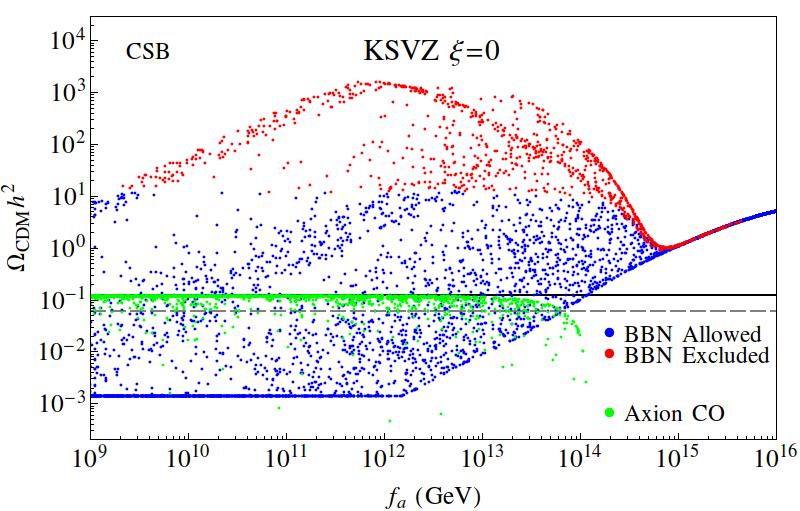}
\caption{
The wino-like WIMP (blue) and axion  (green) relic densities from a scan over
SUSY KSVZ parameter space for the CSB benchmark case with $\xi=0$.
The grey dashed line shows the points where DM consists of 50\% axions and
50\% neutralinos.
\label{fig:csb_ksvz0}}
\end{center}
\end{figure}

In this section, we will examine the CSB benchmark in the SUSY KSVZ case. 
We start with the case where saxion decays into axinos and axions are turned
off ( $\xi=0$).
Results for this benchmark are shown in Fig.~\ref{fig:csb_ksvz0}, where we plot 
$\Omega_{\tz_1}h^2$ (blue points) and $\Omega_a h^2$ (green points)
vs. $f_a$ for the scan over parameters defined in Eq.~(\ref{eq:scan}). 
In the figure, red points violate BBN bounds on late-decaying neutral relics~\cite{jedamzik}
while otherwise the points are BBN safe. 
We also show the measured abundance of CDM by the solid horizontal line. 
Points above this line are excluded by overproduction of dark matter while 
points below the line are allowed.
The dashed horizontal grey line denotes the 50\% CDM abundance so that blue points
above this line have WIMP-dominated CDM while green points above this line have
axion-dominated CDM.

In Fig.~\ref{fig:csb_ksvz0}, one can see that there are three branches of the
neutralino CDM density for $f_a\lesssim10^{15}$ GeV.
These branches reflect three regions of axino mass.
The uppermost branch corresponds to the case of $m_{\ta}<m_{\tz_2}$.
In this case, axinos decay only into $\tz_1$ plus SM particles.
Since the axion sector does not have a direct coupling to a SU(2)$_L$ gauge
supermultiplet, axino decays into $\tz_1$ (mostly wino-like) happen only
through the bino-wino mixing, which is very tiny in the MSSM.
Therefore, for this branch the axino decay occurs well-after the neutralino freeze-out,
enhancing the neutralino abundance well above the measured CDM density
for all values of $f_a$.
Moreover-- for $f_a\gtrsim10^{10}$ GeV-- BBN constrains the model due to the long-lived axino.

The middle branch corresponds to $m_{\tz_2}<m_{\ta}<m_{\tg}$.
In this region axinos can decay directly into $\tz_2$.
Since $\tz_2$ is mostly bino-like and axinos directly couple to
$\tilde{B}$ through the U(1)$_Y$ anomalous coupling, their life-time is much
shorter than in the $m_{\ta} < m_{\tz_2}$ case.
Although the axinos decay after neutralino freeze-out for all $f_a$,
the neutralino density still is smaller than the observed CDM density for
$f_a\lesssim 5\times10^{10}$ GeV.
Hence, both axion-dominated or neutralino-dominated dark matter scenarios
are possible in this region.
For $f_a\gtrsim 5\times 10^{12}$ GeV, all points in the $m_{\ta}<m_{\tg}$ branch
are excluded by BBN.

The lowermost branch corresponds to $m_{\ta}>m_{\tg}$.
In this region, axinos can decay to gluinos through the SU(3)$_c$ anomaly
coupling so that the axino life-time becomes much shorter than the previous two cases.
For $f_a\lesssim 10^{12}$ GeV, axinos decay before neutralino freeze-out in the bulk of this 
parameter region, so the neutralino CDM density takes its standard thermal value $\sim 0.002$.
In the case where the axino mass is close to the gluino mass, however, axinos
can decay after neutralino freeze-out and augment the WIMP abundance.
As $f_a$ increases, axinos more often decay after freeze-out and hence increasingly 
augment the  neutralino relic density.
By $f_a\sim 2\times 10^{12}$ GeV, axinos always decay after freeze-out and always
augment the neutralino abundance. Despite the enhancement of the neutralino
abundance, there are points where the DM is axion-dominated up to $f_a\simeq
6\times 10^{13}$ GeV.

For $f_a\agt 10^{15}$ GeV, the contribution to the WIMP abundance is mostly
from CO-produced saxion decays; these augment the abundance for larger $f_a$
since the saxion CO production rate increases with $f_a$.
On the other hand, the contribution from TP axinos is highly suppressed
for large $f_a$, since the axino thermal production decreases with $f_a$.
Once the TP axino abundance becomes negligible ($f_a \gtrsim 5\times10^{14}$
GeV), the LSP relic abundance becomes independent of the axino mass and all the
branches discussed above collapse into a single line, as seen in
Fig.~\ref{fig:csb_ksvz0}.

\subsubsection{$\xi =1$ case}

\begin{figure}
\begin{center}
\includegraphics[height=6.9cm]{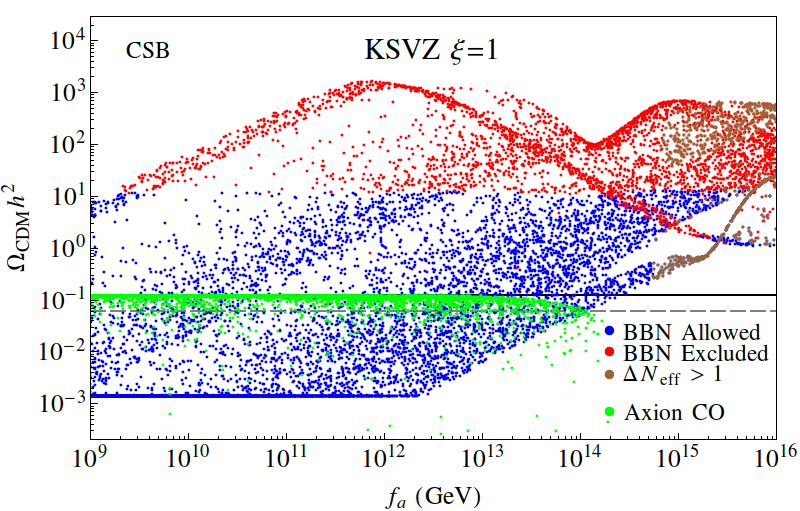}
\caption{
The wino-like WIMP and axion  relic densities from a scan over
SUSY KSVZ parameter space for the CSB benchmark case with $\xi=1$.
The grey dashed line shows the points where DM consists of 50\% axions and
50\% neutralinos.
\label{fig:csb_ksvz1}}
\end{center}
\end{figure}

In Fig.~\ref{fig:csb_ksvz1}, we show $\Omega_{\tz_1,a}h^2$ vs. $f_a$  for the same 
CSB benchmark point but now where saxion decays into axinos and axions are allowed: $\xi=1$. 
For the lower ($f_a\alt 10^{14}$ GeV) range, saxion decays have a smaller
impact on the neutralino abundance and the results are similar to the CSB/KSVZ $\xi=0$ case.
For higher $f_a$ values, CO-produced  saxions become important and since
$s\to aa$ and $\ta\ta$ decays are now allowed, there is a large injection of
relativistic axions. For $f_a>4\times 10^{14}$ GeV, we see brown points which
produce too much dark radiation-- $\Delta N_{\rm eff}>1$-- and are excluded.
There is also a broad band of blue (BBN-allowed) and red (BBN-excluded) points at large
$f_a\sim 10^{15}$ GeV with very high $\Omega_{\tz_1}h^2\sim 1-100$ where the
additional neutralino abundance arises from $s\to\ta\ta$ decays. The lower
disjoint narrow band at $f_a\agt 10^{14}$ GeV and $\Omega_{\tz_1}h^2\sim 0.1-1$
occurs for points where $m_{\ta}>m_s/2$, so $s\to \ta\ta$ is kinematically
forbidden.

Finally we point out that, unlike the $\xi=0$ case, the extremely
large $f_a$ region ($f_a \gtrsim 10^{15}$ GeV) still shows a dependence on the axino mass: this is
responsible for the distinct branches. Although thermal production of axinos is
neglegible in this regime, axinos are non-thermally produced from saxion
decays and can influence the final neutralino abundance.

\subsection{CSB benchmark in SUSY DFSZ}

\subsubsection{$\xi=0$ case}

\begin{figure}
\begin{center}
\includegraphics[height=6.9cm]{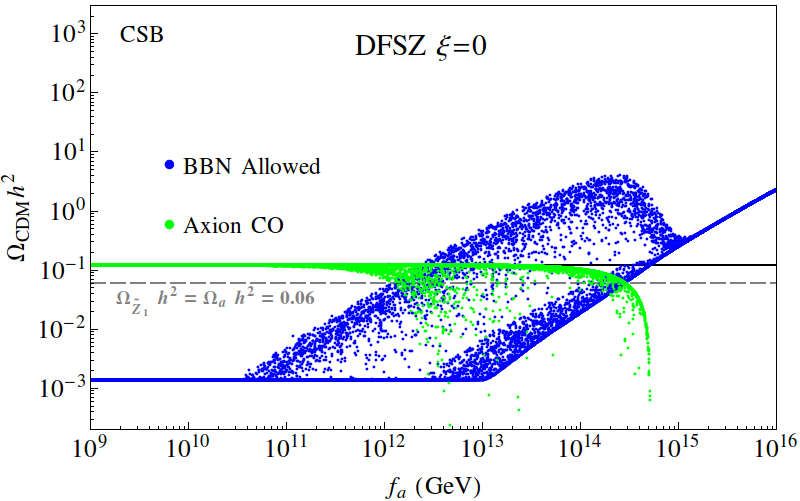}
\caption{
The wino-like WIMP and axion  relic densities from a scan over
SUSY DFSZ parameter space for the CSB benchmark case with $\xi=0$.
The grey dashed line shows the points where DM consists of 50\% axions and
50\% neutralinos.
\label{fig:csb_dfsz0}}
\end{center}
\end{figure}

In this section, we will examine the CSB benchmark in the SUSY DFSZ case. 
As before, we start with the $\xi =0$ case, shown in Fig.~\ref{fig:csb_dfsz0}.
The first noteworthy point is that the large $\mu$ value enhances the
saxion (axino) decay rate to Higgs (higgsinos). As a result the saxion and axino
lifetimes are suppressed and the entire $f_a$ range is BBN safe.
Unlike the KSVZ case, there are two branches for neutralino CDM density, since
in the DFSZ case, the axino decay is determined by the $\mu$-term interaction.
The upper branch corresponds to $m_{\ta}<m_{\tz_3}\sim\mu$ with higgsino-like $\tz_3$.
The axino decay into $\tz_1$ or $\tz_2$ can be through wino-higgsino or bino-higgsino mixing, 
so it is normally suppressed by $(m_Z/\mu)^2$. 
For $f_a\lesssim 3\times10^{10}$ GeV, axinos decay before neutralino freeze-out, 
and thus the neutralino density takes its standard value.
For $f_a\gtrsim 3\times10^{10}$ GeV, axinos tends to decay after neutralino freeze-out 
so the neutralino density gradually increases as $f_a$ increases.
In most of parameter space, axions constitute the bulk of dark matter, 
but wino-like neutralinos can be the dominant dark matter in the region of 
$10^{12}$ GeV$\lesssim f_a\lesssim 10^{13}$ GeV.
By $f_a\gtrsim 10^{13}-10^{14}$ GeV, the neutralino density is typically 
larger than the measured CDM result so the parameter/model choices would be excluded.

The lower branch corresponds to $m_{\ta}>m_{\tz_3}$.
Due to its large interaction, the axino tends to decay before neutralino freeze-out for 
$f_a\lesssim 3\times10^{12}$ GeV.
Therefore, the neutralino relic abundance is usually fixed at its thermally-produced value for
much of the lower range of $f_a$. Once $f_a \gtrsim 10^{13}$ GeV, the neutralino
abundance is always enhanced due to decays of axinos and saxions. Still, the CDM
abundance tends to be axion-dominated for $f_a\alt 2\times 10^{14}$ GeV. For
higher $f_a$ there is a short interval where wino-like WIMPs can dominate the DM abundance.
Finally, for $f_a\agt 6\times 10^{14}$ GeV, WIMP CDM is always
overproduced.
We also point out that for very large $f_a$ values, as in the KSVZ $\xi = 0$ scenario, 
the thermal production of axinos is neglegible hence the neutralino
relic abundance becomes independent of $f_a$.

\subsubsection{$\xi =1$ case}

\begin{figure}
\begin{center}
\includegraphics[height=6.9cm]{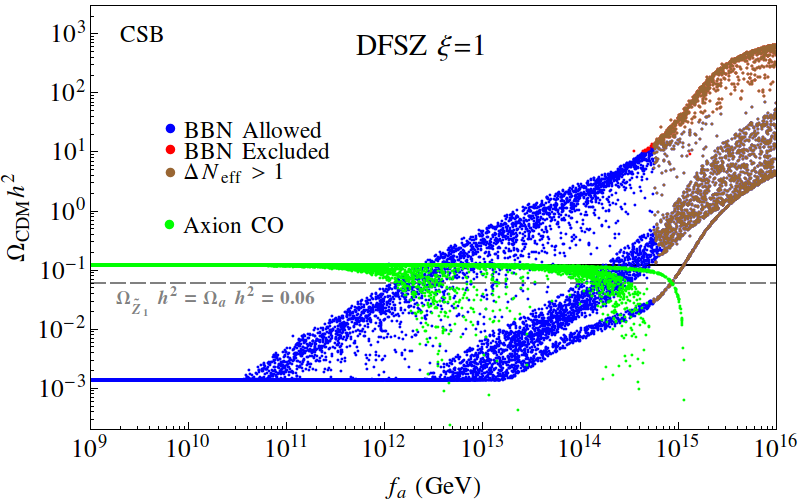}
\caption{
The wino-like WIMP and axion  relic densities from a scan over
SUSY DFSZ parameter space for the CSB benchmark case with $\xi=1$.
The grey dashed line shows the points where DM consists of 50\% axions and
50\% neutralinos.
\label{fig:csb_dfsz1}}
\end{center}
\end{figure}

For the CSB benchmark with SUSY DFSZ and $\xi=1$, the results are shown in 
Fig.~\ref{fig:csb_dfsz1}. The low $f_a$ behavior of the plot is similar to the CSB/DFSZ
case with $\xi =0$: the CDM density is dominated by axions. 
For higher $f_a$ values, where CO-produced saxions become important, 
the saxion lifetime is shortened by the additional contributions from
$s\to aa,\ta\ta$ decays. However, most of the points for $f_a\agt 5\times 10^{14}$ GeV are
forbidden due to overproduction of dark radiation. The lower blue-brown band at 
$f_a\sim 10^{14}-10^{15}$ GeV occurs when $m_{\ta}> m_s/2$ so that additional WIMP
production from $s\to\ta\ta$ is dis-allowed.

\subsection{RNSw benchmark in SUSY KSVZ}

In this subsection we examine dark matter production in the SUSY RNSw benchmark
case.
The RNSw benchmark model has values $m_s=m_0\equiv m_{3/2}=5$ TeV which is far smaller
than that of the CSB benchmark so that saxions (and also axinos since we take their mass to be 
bounded by $m_{3/2}$) are typically much longer-lived than in the CSB case.

\subsubsection{$\xi=0$ case}

\begin{figure}
\begin{center}
\includegraphics[height=6.9cm]{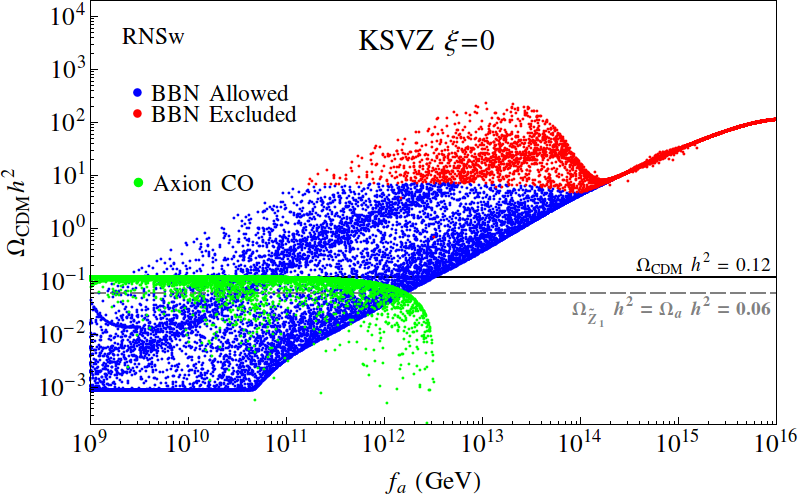}
\caption{
The wino-like WIMP and axion  relic densities from a scan over
SUSY KSVZ parameter space for the RNSw benchmark case with $\xi=0$.
The grey dashed line shows the points where DM consists of 50\% axions and
50\% neutralinos.
\label{fig:ns_ksvz0}}
\end{center}
\end{figure}

In Fig.~\ref{fig:ns_ksvz0}, we plot $\Omega_{\tz_1,a} h^2$ vs. $f_a$ for the SUSY KSVZ case 
using the RNSw benchmark point with $\xi =0$.
In this case, $m_{\ta}$ is always larger than $m_{\tz_2}$, so there are only two branches for the 
neutralino density: $m_{\ta}<m_{\tg}$ and $m_{\ta}>m_{\tg}$.
Long-lived axinos are already augmenting the neutralino relic density 
even at $f_a$ values as low as $10^9$ GeV. 
As we move to higher $f_a$ values, the axinos and saxions are longer-lived, thus contributing even more
to the WIMP abundance. For $f_a$ values $\agt 5\times 10^{12}$ GeV, the model is already excluded
due to overproduction of WIMPs. The $\Omega_{\tz_1}h^2$ points reach even higher values as 
$f_a$ increases until $f_a\sim 3\times 10^{13}$ GeV.
For $3\times 10^{13}$ GeV$\lesssim f_a\lesssim 2\times 10^{14}$ GeV, the axino
contribution decreases due to suppression of the thermal production.
For $f_a \gtrsim 2\times 10^{14}$ GeV, then CO-produced saxions decay into gluino pairs 
and tend to augment the WIMP abundance.
However, this is inconsequential since the model already overproduces WIMP dark matter.
A large BBN forbidden region occurs, but it is already in the WIMP-overproduction region 
so adds no further constraints.

\subsubsection{$\xi =1$ case}

\begin{figure}
\begin{center}
\includegraphics[height=6.9cm]{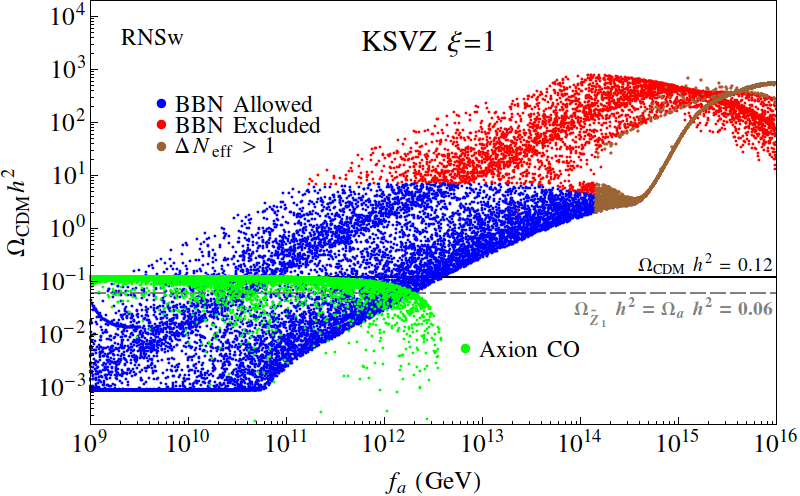}
\caption{
The wino-like WIMP and axion  relic densities from a scan over
SUSY KSVZ parameter space for the RNSw benchmark case with $\xi=1$.
The grey dashed line shows the points where DM consists of 50\% axions and
50\% neutralinos.
\label{fig:ns_ksvz1}}
\end{center}
\end{figure}

For the RNSw benchmark case in SUSY KSVZ with $\xi=1$, as shown in Fig.~\ref{fig:ns_ksvz1},
the low $f_a$ behavior of $\Omega_{\tz_1}h^2$ is very similar to the $\xi= 0$
case, since at low $f_a$ saxion production is not very relevant and saxions
decay well-before neutralino freeze-out. 
For $f_a\agt 4\times 10^{12}$ GeV, as in the $\xi =0$ case, the model over-produces WIMPs
and is excluded. At even larger values of $f_a$, in the DM-excluded region, the $\xi =1$ case
begins to differ from $\xi=0$. An additional branch of $\Omega_{\tz_1}h^2$ appears: the lowermost branch
swings downward due to the suppressed axino and saxion TP as in the $\xi=0$ case, 
but never reaches the DM-allowed line. It is nonetheless
also excluded by overproduction of dark radiation. The upper (also excluded) two branches occur where
$m_s>2m_{\ta}$ so that CO-production of saxions keeps increasing the WIMP abundance.
This region is also excluded by the BBN constraint from late-decaying
saxions followed by axino cascade decays.

\subsection{RNSw benchmark in SUSY DFSZ}

In this subsection, we examine the RNSw benchmark model in the SUSY DFSZ model.
Since $\mu ({\rm RNSw})\ll \mu ({\rm CSB})$, saxions in the RNSw/DFSZ case 
will tend to be longer lived.

\subsubsection{$\xi=0$ case}

\begin{figure}
\begin{center}
\includegraphics[height=6.9cm]{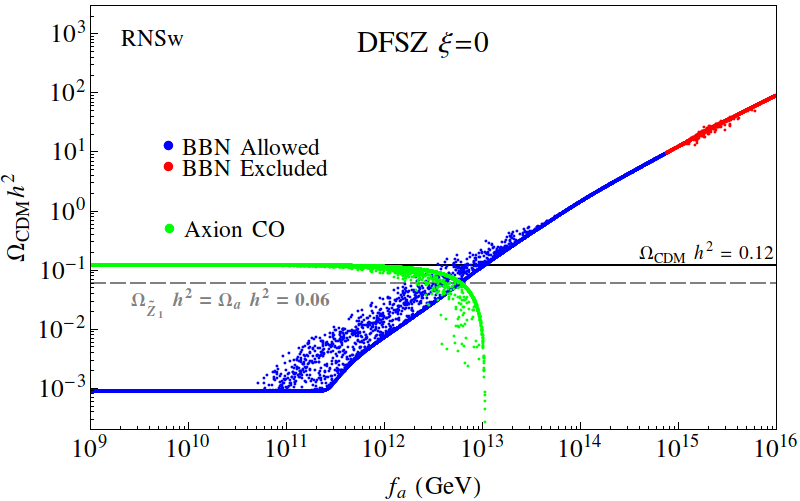}
\caption{
The wino-like WIMP and axion  relic densities from a scan over
SUSY DFSZ parameter space for the RNSw benchmark case with $\xi=0$.
The grey dashed line shows the points where DM consists of 50\% axions and
50\% neutralinos.
\label{fig:ns_dfsz0}}
\end{center}
\end{figure}

In Fig.~\ref{fig:ns_dfsz0} we show RNSw in the SUSY DFSZ case with $\xi =0$. 
In this case, $m_{\ta}$ is always larger than $\mu$, so there is no region corresponding to the upper 
branch in Fig.~\ref{fig:csb_dfsz0}.
For low $f_a$, in contrast to RNSw in the SUSY KSVZ case, the axino lifetime is
smaller and the WIMP abundance remains at its thermally-produced value for
$f_a\alt 5\times 10^{10}$ GeV. 
For higher $f_a$ values, the WIMP abundance is augmented by axino and saxion decays after freeze-out.
Ultimately, the model over-produces WIMPs for $f_a\agt 10^{13}$ GeV. The model tends to be
axion-dominated for $f_a\alt 6\times 10^{12}$ GeV and WIMP dominated for a narrow range
of $f_a$ just beyond this value until WIMP overproduction is reached and the model becomes excluded.
This is in contrast to the CSB benchmark with DFSZ and $\xi =0$, where the
allowed region extends to $f_a\sim 5\times 10^{14}$ GeV 
since saxions and axinos are shorter-lived due to much larger masses 
and stronger interactions ($\mu ({\rm RNSw})\ll \mu ({\rm CSB})$). 

\subsubsection{$\xi =1$ case}

\begin{figure}
\begin{center}
\includegraphics[height=6.9cm]{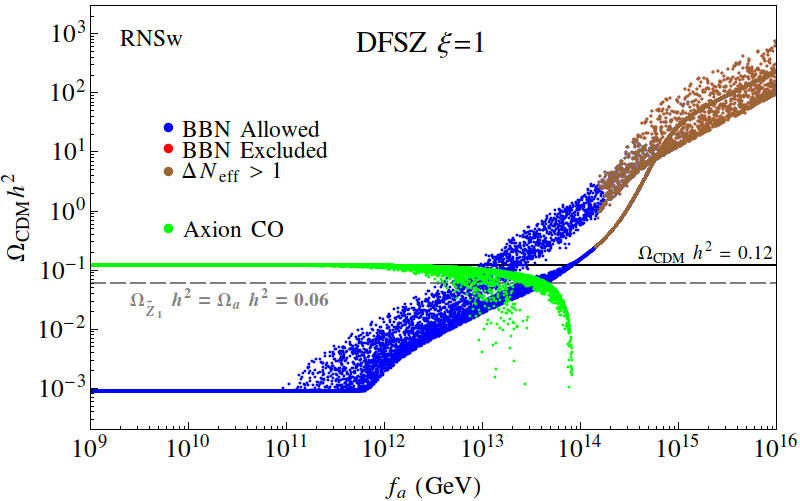}
\caption{
The wino-like WIMP and axion  relic densities from a scan over
SUSY DFSZ parameter space for the RNSw benchmark case with $\xi=1$.
The grey dashed line shows the points where DM consists of 50\% axions and
50\% neutralinos.
\label{fig:ns_dfsz1}}
\end{center}
\end{figure}

In Fig.~\ref{fig:ns_dfsz1} we show results for the RNSw benchmark in SUSY DFSZ with
$\xi =1$. While the low $f_a$ behavior is similar to the results from the $\xi=0$
case, the high $f_a$ behavior is different. The decays $s\to aa$ and $s\to \ta\ta$ 
allow the saxion to decay more quickly than in the $\xi =0$ case for a common value of $f_a$.
Thus, the DM-allowed region extends to larger $f_a$ values: in this case up to
$f_a\sim 10^{14}$ GeV. For these high $f_a$ values, the relic density band again splits into
two branches: one with heavy axinos (lower-branch),
where $s\to \ta\ta$ is closed, and one with light axinos (upper
branch), where $s\to\ta\ta$ is open, thus augmenting the WIMP abundance.
The points with $f_a\agt 2\times 10^{14}$ GeV tend to be doubly-excluded by overproduction
of WIMPs and by overproduction of dark radiation.

\section{Dependence of mixed axion-wino abundance on 
sparticle mass spectra}
\label{sec:lines}

In the previous sections we have investigated the DM-allowed range of 
$f_a$ for two SUSY benchmark models with wino-like LSPs, $f_a$ and $m_{\ta}$ as
free parameters and $m_s=m_0$.
In this section, we investigate how our results might change as a function of
the MSSM spectrum.
To explore this issue, we extend our two benchmark points into model lines in the MSSM sector.
For brevity, we consider here only the DFSZ model-- which provides a solution to the SUSY $\mu$ problem-- 
to see the impacts of axino/saxion production/decays on the CDM density.
Actually, even in the presence of late-decaying axinos and saxions, 
the most important factor that determines the WIMP abundance is the WIMP-WIMP 
annihilation cross section since the augmented density is 
determined mainly by annihilation cross section evaluated at the heavy particle decay temperature: 
this is the case of so-called WIMP re-annihilation after non-thermal WIMP production
from heavy particle decay~\cite{az1}.
For this reason, the behavior of our plots is similar for both DFSZ and KSVZ models, 
and so we will show only the DFSZ case and then briefly comment on the KSVZ case.

\subsection{CSB model line}

For the CSB benchmark, we will now allow $m_{3/2}$ to vary while
keeping $\tan\beta$ fixed at 10 with $\mu =3$ TeV and $m_A=m_{3/2}$.
For the CSB model-line, we require $m_{3/2}\agt 32$ TeV so the
mass of the lightest wino-like chargino is always above the limit
$m_{\tw_1}\agt 91.9$ GeV established from LEP2 searches. 
The upper limit on $m_{3/2}$ occurs at $\sim 115$ TeV where the 
predicted value of $m_h$ climbs above $128$ GeV. Here, we allow for 
an expected theory error in the Isasugra calculation of 
$m_h$ at about $\pm 2.5$ GeV.

\begin{figure}
\begin{center}
\includegraphics[height=6.9cm]{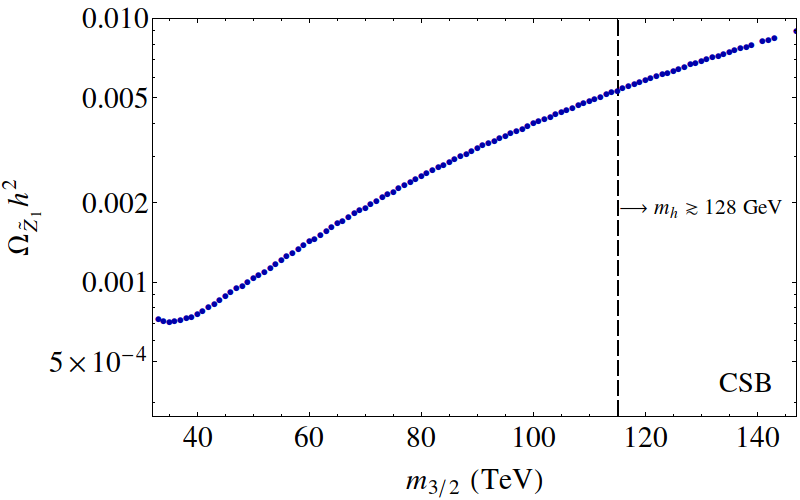}
\caption{Plot of thermally-produced neutralino abundance 
$\Omega_{\tz_1}h^2$ vs. $m_{3/2}$ along the CSB model line with
$\tan\beta =10$ and $\mu =3$ TeV.
\label{fig:csb_om}}
\end{center}
\end{figure}
We show the thermally-produced neutralino abundance for the CSB model line
in Fig.~\ref{fig:csb_om}. Here, we see that $\Omega_{\tz_1}^{\rm TP}h^2$
ranges from around $0.0007$ at the lower limit to about $0.005$
at the upper limit as compared to $0.002$ for the CSB benchmark.
Roughly speaking, the thermally-produced wino abundance will provide either
more or less room in general for non-thermally produced winos and axions.

\begin{figure}
\begin{center}
\includegraphics[height=6.9cm]{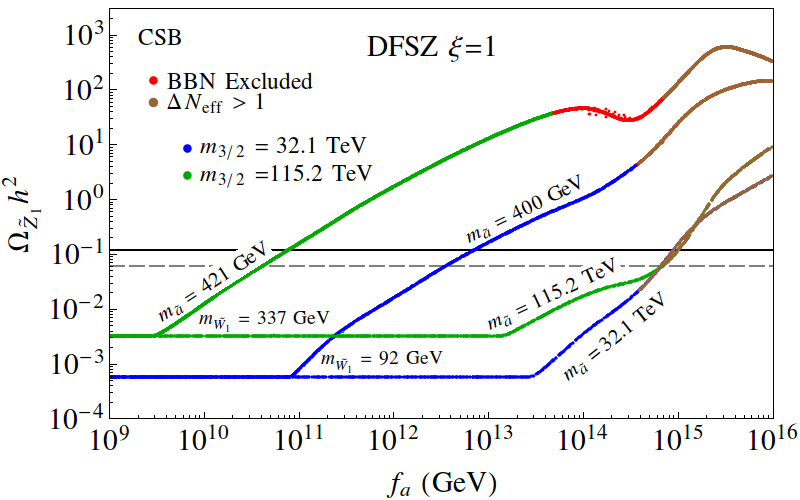}
\caption{Plot of thermally- and non-thermally-produced neutralino abundance 
$\Omega_{\tz_1}h^2$ vs. $f_a$ along the CSB model line in the DFSZ $\xi=1$ 
case for a light 
($m_{3/2}=32$ TeV, blue envelope) and heavy ($m_{3/2}=115$ TeV, green envelope) 
CSB mass spectrum where $m_{\ta}$ ranges from 400 GeV up to $m_{3/2}$.
\label{fig:csb_sh}}
\end{center}
\end{figure}
In Fig.~\ref{fig:csb_sh}, we show the value of $\Omega_{\tz_1}h^2$ which 
is produced from the coupled Boltzmann calculation of mixed axion-wino CDM
versus $f_a$ for the minimal and maximal values of $m_{3/2}$ which are
allowed along the CSB model line. The blue curves provide the calculated 
envelope of values for the lower limit of $m_{3/2}\sim 32$ TeV. 
At low $f_a$, $\Omega_{\tz_1}h^2$ lies at the TP-value $\sim 0.0006$ since 
thermally-produced axinos always decay before neutralino freeze-out. 
As $f_a$ climbs above $\sim 10^{11}$ GeV, then the lighter axinos start 
decaying after neutralino freeze-out whilst the heavier axinos still decay before freeze-out. 
The region between the two blue curves shows the range of
$\Omega_{\tz_1}h^2$ which is generated for $0.4$ TeV $<m_{\ta}<$ 32 TeV.
We see that values of $f_a$ up to $\sim 10^{15}$ GeV are dark-matter-allowed
for very heavy axinos. However, at values of $f_a\agt 5\times 10^{14}$ GeV, 
then too much dark radiation is produced from $s\to aa$ decays in addition to WIMP overproduction so 
that the parameter space is doubly-excluded.

The heavy end of the CSB model line $m_{3/2}=115$ TeV is shown by the
envelope of green curves. 
For the light axino with $m_{\ta}=421$ GeV, the thermally-produced value of 
$\Omega_{\tz_1}h^2\sim 0.003$ is obtained only for the short range 
of $f_a\alt 3\times 10^9$ GeV. For this heavy CSB spectra, the gauginos are all 
sufficiently heavy and thus the axino can decay only into $\tz_1$ so that the axino lifetimes are 
much longer than that in the case for light spectra. 
The upper range of the green envelope comes from light axino masses where $m_{\ta}=421$ GeV
is the threshold for $\ta\to Z\tz_1$ decay hence augmenting neutralino density at low $f_a$,
while the lower envelope is established by the heaviest axino mass values.
For the upper part of the envelope, the red points denote the on-set of 
BBN bounds on late decaying saxions as ruling out $f_a\agt 5\times 10^{13}$ GeV.
For the lower part of the envelope, with axino masses ranging to
32 TeV, then values of $f_a$ up to $5\times 10^{14}$ GeV are possible.


In the case of KSVZ model, the spectrum dependence is similar to the DFSZ model.
For the light spectum ($m_{3/2}=32$ TeV), the neutralino abundance tends to be smaller 
due to its large annihilation cross section.
For the heavy spectrum ($m_{3/2}=115$ TeV), the cross section becomes larger, 
so the neutralino abundance becomes smaller.
Nevertheless, the allowed range of $f_a$ for the heavy spectrum is slightly larger
than that for light spectrum since the saxion mass is larger
($m_s=m_{3/2}$) so that its decay can occur earlier.

\subsection{RNSw model line}

For the RNSw benchmark, we will instead allow the GUT scale 
SU(2)$_L$ gaugino mass $M_2$ to vary while
keeping $m_0=5$ TeV, $m_{1/2}=700$ GeV, $A_0=-8$ TeV and 
$\tan\beta$ fixed at 10 with $\mu =200$ GeV and $m_A=1$ TeV.
For the RNSw model-line, the lower limit on $M_2$ is again set by 
the limit from LEP2 searches for wino-like charginos. The upper limit
on $M_2\alt 250$ GeV is set from simply requiring a wino-like LSP:
for higher $M_2$ values, the lightest neutralino becomes
increasingly higgsino-like, a case which was shown in Ref's~\cite{dfsz1, Bae:2014rfa}.
The naturalness value $\Delta_{EW}$ remains fixed at around 10 
since varying $M_2$ hardly affects it~\cite{bbhmpt}.

\begin{figure}
\begin{center}
\includegraphics[height=6.9cm]{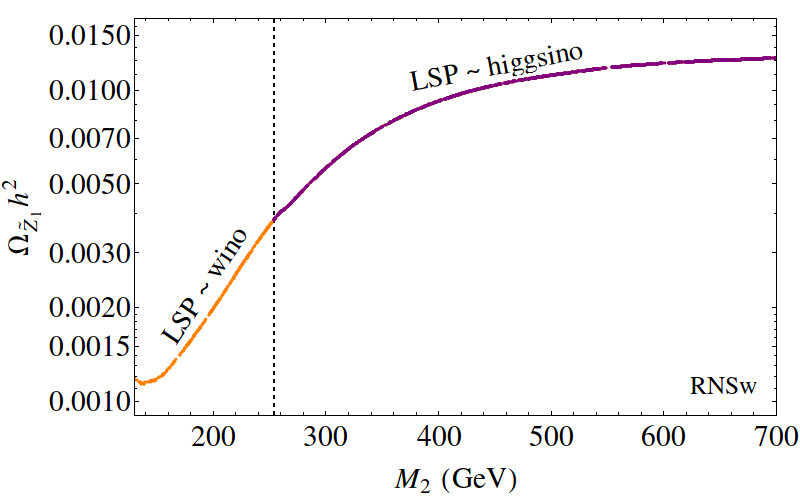}
\caption{Plot of thermally-produced neutralino abundance 
$\Omega_{\tz_1}h^2$ vs. $M_2$ along the RNSw model line.
\label{fig:rns_om}}
\end{center}
\end{figure}
\begin{figure}
\begin{center}
\includegraphics[height=6.9cm]{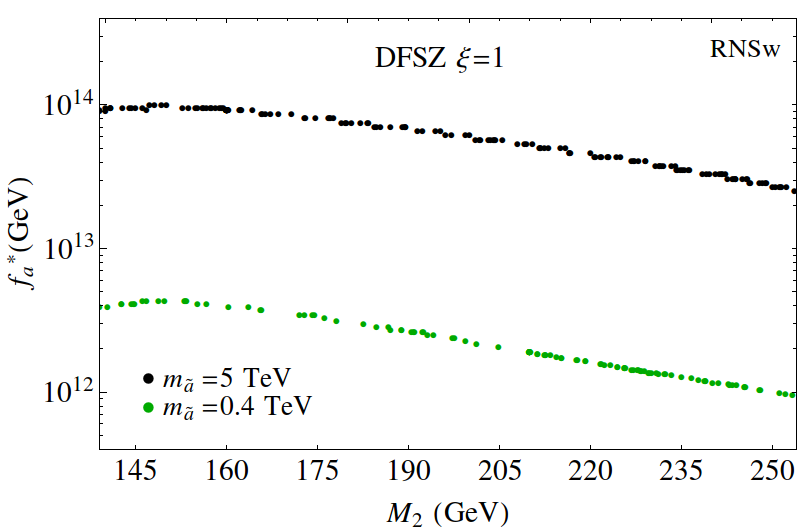}
\caption{Plot of upper limit of $f_a$ allowed from the
RNSw model line in the DFSZ case with $\xi =1$ 
versus $M_2$ for $m_{\ta}=400$ GeV and $m_{\ta}=5$ TeV.
\label{fig:rns_fa}}
\end{center}
\end{figure}

We show the thermally-produced neutralino abundance for the RNSw model line
in Fig.~\ref{fig:rns_om}. The lower range of
$\Omega_{\tz_1}^{\rm TP}h^2$ occurs at $0.0012$ for $M_2\sim 140$ GeV. 
The maximal value reaches up to $\sim 0.004$ before entering the 
higgsino-like LSP region. Since the thermally-produced wino abundance
increases with $M_2$, the allowed enhancement from non-thermal production
decreases as $M_2$ increases. Furthermore, since the non-thermal production
(from saxion and axino decays) grows with $f_a$, we expect the maximum allowed
value for $f_a$ to decrease as $M_2$ increases.
This is shown in Fig.~\ref{fig:rns_fa}, where we plot the upper limit on $f_a$,
denoted as $f_a^*$, versus $M_2$ along the RNSw model line in the DFSZ $\xi=1$ case. 
The axino mass is $m_{\ta}=0.4$ TeV (green dots) or $m_{\ta}=5$ TeV (black dots).
The upper limit comes only from the overproduction of WIMPs in this case 
since violation of BBN bounds and overproduction of dark matter occurs at higher $f_a$ in this model.

In the KSVZ model, on the other hand, axinos are longer-lived than in the DFSZ
model, so the allowed range of $f_a$ is smaller than that in the DFSZ case.
As we have seen in Fig.~\ref{fig:ns_ksvz1}, only a small region,
$f_a\lesssim{\cal O}(10^{10})$ GeV, is allowed for $m_{\ta}\lesssim m_{\tg}$,
while $f_a\lesssim{\cal O}(10^{12})$ GeV is allowed for $m_{\ta}=5$ TeV.

\section{Conclusion}
\label{sec:conclude}

In this paper we have examined mixed axion/wino cold dark matter production in 
two SUSY benchmark models with a wino as LSP. 
The first-- labeled as CSB-- is typical of a variety of 
models (PeV-SUSY, some split SUSY variations, KL, PGM, spread SUSY) with a thermally-underproduced
wino-like WIMP abundance. The second, labeled as RNSw, is a model with radiatvely-driven naturalness
but with a wino-like rather than a higgsino-like  LSP. Our calculation of mixed axion/wino
dark matter production stands in contrast to the more commonly examined case of non-thermal
WIMP production due to late decaying moduli fields~\cite{mr,gondolo,g2DM}. We find it a more
appealing method for augmenting the dark matter abundance since it also provides a solution
to the strong CP problem and-- in the case of SUSY DFSZ-- provides for a solution to the
SUSY $\mu$ problem. 

We have presented results for the wino-like
WIMP abundance and axion abundance as a function of the axion decay constant $f_a$
and the axino mass $m_{\ta}$. In the bulk of the parameter space, 
WIMPs are thermally under-produced at low and intermediate $f_a$ values
($\sim 10^9-10^{11}$ GeV) so that the DM abundance tends to be axion-dominated.
This has important consequences for direct and indirect WIMP detection experiments 
since it anticipates a greatly reduced local abundance of WIMPs and hence
diminished prospects for wino-like WIMP detection. This can actually
allow for wino-like WIMP dark matter to evade the recent Fermi~\cite{fermi} 
searches for gamma ray emission from dwarf-spheroidal galaxies since in this case
the expected event rate is expected to be reduced by a factor $(\Omega_{\tw}h^2/0.12)^2$.

\begin{figure}
\begin{center}
\includegraphics[height=6.9cm]{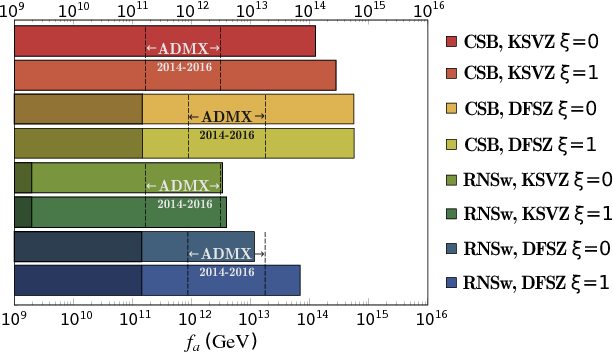}
\caption{Range of $f_a$ which is allowed in each PQMSSM scenario for the
CSB and RNSw benchmark models. 
Darker-shaded regions indicate the range of $f_a$ where $\theta_i > 3$ which
might be considered unnatural. We also show the $f_a$ range which is expected to
be probed by the ADMX experiment in the next few years.
\label{fig:bar}}
\end{center}
\end{figure}
A grand overview of our results is presented in Fig.~\ref{fig:bar}, where we
show the allowed range of $f_a$ as a bar for each of the two benchmark points
and for each of the eight SUSY PQ models considered. 
For all the models, no GUT scale values of $f_a$ ($f_a \sim 10^{16}$ GeV)
are allowed. This is due to the rather large value of
$m_s\sim m_{3/2}$ in our benchmark models.
In these cases, saxions always decay to SUSY particles
and no entropy dilution of WIMPS and axions is
possible (see Refs.~\cite{dfsz1,Bae:2014rfa} for more details).

In addition to the SUSY KSVZ case with SU(2)$_L$ singlet heavy quark states 
which has been presented here, 
we have also investigated SUSY KSVZ models including SU(2)$_L$ doublet heavy quark states 
so that the axion superfield has couplings with SU(2)$_L$ gauge superfields.
In the case of doublet heavy quarks, axino decays to the
wino-like neutralino are not suppressed, even for $m_{\ta}<m_{\tz_2}$.
Therefore, there is no separate branch like the uppermost one in 
Fig.~\ref{fig:csb_ksvz0} and \ref{fig:csb_ksvz1} and thus there are only two
branches determined by $m_{\tg}$.
The basic features of plots with doublet heavy quarks are similar to the case 
with singlet heavy quarks since the
dominant axino decay mode is into gluinos for both cases.
The allowed range of $f_a$ values is extended only slightly for doublet KSVZ heavy
quarks as compared to the case of singlet heavy quarks shown in this paper.

For sufficiently heavy axinos, all models shown in this paper are DM-allowed for the lower
range of $f_a\sim 10^9-10^{12}$ GeV, since WIMPs are underproduced. 
In these cases, the remaining abundance is made up of axions. 
Even though one might expect a low axion abundance at low $f_a$
in the case where the initial mis-alignment angle is $\theta_i\sim \mathcal{O}(1)$, 
due to anharmonicity effects the necessary
axion abundance can always be obtained by taking $\theta_i\sim \pi$~\cite{vg}. 
In this case, one might wonder about 
fine-tuning of the axion abundance such that the axion fields sits atop the peak of
its potential. 
Thus, for cases where $\theta_i > 3$, we shade these regions as darker in Fig.~\ref{fig:bar}. 
The non-shaded regions may be more natural as far as the expected initial axion field value goes. 

We should note that for KSVZ models, regions with $\theta_i < 3$ at low $f_a$ occur only if 
the wino LSP constitutes more than $\sim90\%$ of total CDM density since 
axion CO-production is very low for $f_a \alt 10^{10}$.
Then, the CSB benchmark in the SUSY KSVZ model most naturally allows for the lowest
$f_a$ values while the CSB benchmark in the DFSZ model allows for the highest $f_a$ values.
The range of $f_a$ values obtained for the RNSw benchmark is more constrained than the CSB case.
The upper bounds on $f_a$ for the two benchmark models are well-maintained even when the points are
extended to model lines, as was shown for the DFSZ $\xi=1$ case in Sec.~\ref{sec:lines}.

Finally, we denote the range of $f_a$ values which are expected to be probed in the next few
years by the ADMX experiment~\cite{admx}. The values shift between KSVZ and DFSZ models 
since the domain wall number $N_{\rm DW}=1$ for KSVZ and 6 for DFSZ and $m_a\simeq 0.62\ {\rm eV}[ 10^7\ {\rm GeV}/(f_a/N_{\rm DW}) ]$.
We also note that a possible ADMX technique of open resonators~\cite{Rybka:2014cya} may allow
even lower values of $f_a$ to be probed in the future.

\acknowledgments

We thank the William I. Fine Theoretical Physics Institute (FTPI) at the University of 
Minnesota for hospitality while the bulk of this work was completed.
The computing for this project was performed at the 
OU Supercomputing Center for Education \& Research (OSCER) at the University of
Oklahoma (OU). A. L. thanks Fundac\~ao de Amparo \`a Pesquisa do Estado de
S\~ao Paulo (FAPESP) for supporting this work.

\end{document}